\documentclass[preprint,fleqn,showpacs,showkeys]{revtex4}
\usepackage{graphicx}
\usepackage{amssymb}
\usepackage{amsmath}
\usepackage{bm}
\begin{document}
\setcounter{page}{0}
\title[]{Charmonium dissociation in collision with $\phi$ meson\\
in hadronic matter}
\author{Shi-Tao \surname{Ji}}
\email{459756153@qq.com}
\thanks{Fax: +86 21 66134208}
\author{Xiao-Ming \surname{Xu}}
\affiliation{Department of Physics, Shanghai University, Baoshan,
 Shanghai 200444, China}

\begin{abstract}
The $\phi$-charmonium dissociation reactions in hadronic matter are studied.
Unpolarised cross sections for
$\phi J/\psi\to D^-_s D^+_s$,
$\phi J/\psi\to D^{*-}_s D^+_s$ or $D^-_s D^{*+}_s$,
$\phi J/\psi\to D^{*-}_s D^{*+}_s$,
$\phi \psi'\to D^-_s D^+_s$,
$\phi \psi'\to D^{*-}_s D^+_s$ or $D^-_s D^{*+}_s$,
$\phi \psi'\to D^{*-}_s D^{*+}_s$,
$\phi \chi_c\to D^-_s D^+_s$,
$\phi \chi_c\to D^{*-}_s D^+_s$ or $D^-_s D^{*+}_s$
and $\phi \chi_c\to D^{*-}_s D^{*+}_s$ are calculated in the Born
approximation, in the quark-interchange mechanism and with a
temperature-dependent quark potential. The potential leads to remarkable 
temperature dependence of the cross sections. With the cross sections
and the $\phi$ distribution function we calculate the
dissociation rates of the charmonia in the interactions with the $\phi$ meson
in hadronic matter. The dependence of the rates on temperature and charmonium 
momentum is meaningful to the influence of $\phi$ 
mesons on charmonium suppression.
\end{abstract}



\maketitle

\section{INTRODUCTION}
Hadronic matter affects production of particles in relativistic heavy-ion collisions.
Change of $J/\psi$ in hadronic matter is appreciable. The change of $J/\psi$ number
is caused by inelastic
meson-charmonium scattering \cite{PRC98Matinyan,NPB79Peskin,PRC95Martins,PRC12Zhou,JPG15Ji}
and spontaneous dissociation that may happen when temperature is higher than the $J/\psi$ dissociation
temperature \cite{PRC02Wong}.
Studies of charmonium dissociation in collisions with hadrons are important in the field of relativistic
heavy-ion collisions. Since the $\phi$-meson production is enhanced in Au-Au
collisions at Relativistic Heavy Ion Collider energies, it is interesting to
investigate $\phi$-charmonium dissociation
in hadronic matter. This is the subject of this paper.

We calculate $\phi$-charmonium dissociation cross sections in the Born approximation and in the
quark-interchange mechanism \cite{PRD92Barnes}.
The quark interchange mechanism between the incident meson and the charmonium breaks
the charmonium and produces charmed mesons and/or charmed strange mesons. Dissociation
of charmonia in collisions with $\pi$, $\rho$ and $K$ in vacuum has been studied in
Ref. \cite{PRC95Martins} and dissociation in hadronic matter in
Refs. \cite{PRC12Zhou,JPG15Ji}. Meson-charmonium dissociation cross sections in
hadronic matter differ from those in vacuum.
The in-vacuum cross sections obtained in the quark-interchange approach are different
from the cross sections obtained in the meson-exchange approach \cite{PRC98Matinyan} and
in the short-distance approach \cite{NPB79Peskin}.
The $\phi$-charmonium dissociation cross sections have not been calculated in the
quark-interchange approach, in the meson-exchange approach and in the short-distance
approach.

With the $\phi$-charmonium dissociation cross sections we calculate the dissociation
rate of the charmonium in the interaction with the $\phi$ meson in hadronic matter. The larger
the dissociation rate is, the stronger charmonium suppression $\phi$ mesons cause. From
the dissociation rate we can understand contribution of the $\phi$ meson to the
charmonium suppression. Therefore, the second task of the paper is to calculate the dissociation
rate.

The remainder of this paper is organized as follows. In Section
II formulas for the dissociation cross section and the dissociation rate are presented.
In Section III we show numerical results of unpolarised $\phi$-charmonium dissociation cross sections
and charmonium dissociation rates. Relevant discussions are given. A summary is in Section IV.

\section{Cross-section and dissociation-rate formulas}
The scattering $A(s\bar{s})+B(c\bar{c})\to C(s\bar{c})+D(c\bar{s})$ takes place due to
quark interchange and interactions among constituents (quarks and antiquarks). The scattering has the
two forms, the prior form and the post form. The scattering in the prior form
means that gluon exchange occurs before quark interchange. The scattering in the
post form contains gluon exchange after quark interchange. The scattering in the
two forms is depicted in Figs. 1 and 2 in Ref. \cite{JPG15Ji}. Cross-section formulas
for the scattering are given in Refs. \cite{PRC12Zhou,NPA07Li}. Here we briefly introduce the formulas.
Application of the cross section to get the charmonium dissociation rate follows.

Denote the spin, orbital angular momentum and four-momentum of meson $i$ ($i=s\bar{s}$, $c\bar{c}$,
$s\bar{c}$, $c\bar{s}$) by $S_i$, $L_i$ and $P_i=(E_i,\vec{P}_i)$, respectively. The Mandelstam
variables are $s=(E_{s\bar{s}}+E_{c\bar{c}})^2-(\vec{P}_{s\bar{s}}+\vec{P}_{c\bar{c}})^2$
and $t=(E_{s\bar{s}}-E_{s\bar{c}})^2-(\vec{P}_{s\bar{s}}-\vec{P}_{s\bar{c}})^2$. Let $\vec{P}$
and $\vec{P}'$ be the
momenta of mesons $A$ and $C$ in the center-of-mass frame of $A$ and $B$, respectively.
The unpolarised cross section for the scattering in the prior form is
\begin{eqnarray}
\sigma^{\rm unpol}_{\rm prior} &=& \frac{1}{(2S_{s\bar s} +1) (2S_ {c\bar c}+1) (2L_{c\bar c}+1)}
\frac{1}{32\pi s}\frac{|\vec{P}^{\prime }(\sqrt{s})|}
{|\vec{P}(\sqrt{s})|}\sum \limits_ {SL_{c\bar cz}} (2S+1)\nonumber\\
&&\int_{0}^{\pi }d\theta|\mathcal{M}_{\rm fi}^{\rm prior} (s,t)|^{2}\sin \theta,
\label{eq:sigmaprior2}
\end{eqnarray}
where $\mathcal{M}_{\rm fi}^{\rm prior}$ is the transition amplitude for the scattering in the
prior form, $S$ is the total spin of mesons $A$ and $B$, $L_{c\bar cz}$ is the magnetic projection
quantum number of $L_{c\bar c}$ and $\theta$ is the angle between $\vec{P}$ and $\vec{P}'$.
The unpolarised cross section for the scattering in the post form is
\begin{eqnarray}
\sigma^{\rm unpol}_{\rm post} &=& \frac{1}{(2S_{s\bar s} +1) (2S_ {c\bar c}+1) (2L_{c\bar c}+1)}
\frac{1}{32\pi s}\frac{|\vec{P}^{\prime }(\sqrt{s})|}
{|\vec{P}(\sqrt{s})|}\sum \limits_ {SL_{c\bar cz}} (2S+1)\nonumber\\
&&\int_{0}^{\pi }d\theta|\mathcal{M}_{\rm fi}^{\rm post} (s,t)|^{2}\sin \theta,
\label{eq:sigmapost2}
\end{eqnarray}
where $\mathcal{M}_{\rm fi}^{\rm post}$ is the transition amplitude for the scattering in the
post form. The unpolarised cross section for $A(s\bar{s})+B(c\bar{c})\to C(s\bar{c})+D(c\bar{s})$
is
\begin{eqnarray}
\sigma^{\rm unpol}=\frac{1}{2}(\sigma^{\rm unpol}_{\rm prior}+\sigma^{\rm unpol}_{\rm post}).
\end{eqnarray}

Let $\psi_{s\bar{s}}$ ($\psi_{c\bar{c}}$, $\psi_{s\bar{c}}$, $\psi_{c\bar{s}}$) stand for
the product of colour, spin, flavour and relative-motion wave functions of $s\bar{s}$ ($c\bar{c}$,
$s\bar{c}$, $c\bar{s}$) and $V_{ab}$ the interaction between constituents $a$ and $b$.
The transition amplitudes are given by
\begin{eqnarray}
{\cal M}_{\rm fi}^{\rm prior} =
4\sqrt {E_{s\bar{s}} E_{c\bar{c}}E_{s\bar{c}}E_{c\bar{s}}}
\langle\psi_{s\bar {c}}|\langle\psi_{c\bar{s}}|
(V_{s\bar {c}} +V_ {c \bar {s}} +V_ {sc}+V_{\bar {s}\bar {c}})
|\psi_{s\bar{s}}\rangle|\psi_{c\bar {c}}\rangle, \label{eq:mprior}
\end{eqnarray}
\begin{eqnarray}
{\cal M}_{\rm fi}^{\rm post} =
4\sqrt {E_{s\bar{s}} E_{c\bar{c}}E_{s\bar{c}}E_{c\bar{s}}}
\langle\psi_{s\bar {c}}|\langle\psi_{c\bar{s}}|
(V_{s\bar{s}} +V_ {c \bar {c}} +V_ {sc}+V_{\bar {s}\bar {c}})
|\psi_{s\bar{s}}\rangle|\psi_{c\bar {c}}\rangle. \label{eq:mpost}
\end{eqnarray}

The potential $V_{ab}$ in coordinate space is
\begin{eqnarray}
V_{ab}(\vec {r}) = V_{\rm{si}}(\vec {r})+V_{\rm{ss}}(\vec {r}),
\end{eqnarray}
where $\vec {r}$ is the relative coordinate of $a$ and $b$,
$V_{\rm{si}}$ is the central spin-independent potential and $V_{\rm{ss}}$ the
spin-spin interaction. $V_{\rm{si}}$ is given by
\begin{eqnarray}
V_{\rm {si}}(\vec {r})=
-\frac {\vec {\lambda}_a}{2} \cdot \frac {\vec{\lambda}_b}{2}
\frac{3}{4} D \left[ 1.3- \left( \frac {T}{T_{\rm c}} \right)^4 \right]\tanh (Ar)
+ \frac {\vec {\lambda}_a}{2} \cdot \frac {\vec {\lambda}_b}{2}
\frac {6\pi}{25} \frac {v(\lambda r)}{r} \exp (-Er),
\label{eq:vsi}
\end{eqnarray}
where $D=0. 7$ GeV, $T_{\rm c}=0.175$ GeV, $A=1.5[0.75+0.25
(T/{T_{\rm c}})^{10}]^6$ GeV, $E=0. 6$ GeV and $\lambda=\sqrt{3b_0/16\pi^2 \alpha'}$
in which $\alpha'=1.04$ GeV$^{-2}$ and $b_{0}=11-\frac{2}{3}N_{f}$
with the quark flavour number $N_{f}=4$. $\vec {\lambda}_a \cdot \vec
{\lambda}_b$ is the product of the Gell-Mann matrices for the colour generators
of $a$ and $b$. The dimensionless function $v(x)$ is given
by Buchm\"{u}ller and Tye \cite{PRD81Buchmuller}. The short-distance part of
$V_{\rm si}$ originates from one-gluon exchange plus perturbative one- and two-loop
corrections. The intermediate-distance and large-distance part of $V_{\rm si}$ fits
well the numerical potential obtained in the lattice gauge calculations \cite{NPB01Karsch}.
At large distances $V_{\rm si}$ is independent of $\vec {r}$ and
exhibits a plateau. The plateau height decreases
with increasing temperature. This means that confinement becomes weaker and weaker.

The spin-spin interaction with relativistic effects \cite{PRD92Barnes,PRD85Godfrey}
is \cite{NPA02Xu,PRC12Zhou}
\begin{eqnarray}
V_{\rm ss}(\vec {r})=
- \frac {\vec {\lambda}_a}{2} \cdot \frac {\vec {\lambda}_b}{2}
\frac {16\pi^2}{25}\frac{d^3}{\pi^{3/2}}\exp(-d^2r^2) \frac {\vec {s}_a \cdot \vec
{s} _b} {m_am_b}
+ \frac {\vec {\lambda}_a}{2} \cdot \frac {\vec {\lambda}_b}{2}
  \frac {4\pi}{25} \frac {1} {r}
\frac {d^2v(\lambda r)}{dr^2} \frac {\vec {s}_a \cdot \vec {s}_b}{m_am_b},\label{eq:vss}
\end{eqnarray}
where $\vec {s}_a$($\vec {s}_b$) and $m_a$($m_b$) are the spin and mass of constituent $a$($b$),
respectively. The flavour dependence of the interaction is relevant to quark masses as
shown in $\frac{1}{m_am_b}$ and $d$,
\begin{eqnarray}
d^2=\sigma_{0}^2\left[\frac{1}{2}+\frac{1}{2}\left(\frac{4m_a m_b}{(m_a+m_b)^2}\right)^4\right]
+\sigma_{1}^2\left(\frac{2m_am_b}{m_a+m_b}\right)^2,
\label{eq:d}
\end{eqnarray}
where $\sigma_0=0.15$ GeV and $\sigma_1=0.705$. The second term in Eq.~(\ref{eq:vss}) arises
from the perturbative one- and two-loop corrections to the gluon propagator.

By solving the Schr\"{o}dinger equation with the central spin-independent potential
plus the spin-spin interaction, meson masses and quark-antiquark relative-motion wave functions
are obtained. In solving the Schr\"{o}dinger equation the masses of 
the charm quark, the up
quark and the strange quark are 1.51 GeV, 0.32 GeV and 0.5 GeV, respectively. The experimental masses
of $\phi$, $J/\psi$, $\psi'$, $\chi_c$, $D_s$ and $D^*_s$ mesons \cite{CPC14Olive} and
the experimental data of $S$-wave $I=$ 2 elastic
phase shifts for $\pi\pi$ scattering in vacuum \cite{PRD71Colton}
are reproduced with $V_{ab}(\vec {r})$ at $T=$ 0 and the quark-antiquark relative-motion wave functions.

While $V_{ab}(\vec {r})$ and the quark-antiquark relative-motion wave functions obtained
from the Schr\"{o}dinger equation with the potential are used to calculate ${\cal M}^{\rm prior}_{\rm fi}$
and ${\cal M}^{\rm post}_{\rm fi}$, we get ${\cal M}^{\rm prior}_{\rm fi}=
{\cal M}^{\rm post}_{\rm fi}$. This is exactly what is concluded in Ref. 
\cite{Clarendon65Mott} from a general argument.

With the unpolarised cross sections for charmonium dissociation we calculate
the dissociation rate of charmonium in the interaction with $\phi$ meson in hadronic matter,
\begin{eqnarray}
n\langle v_{\rm rel} \sigma^{\rm unpol} \rangle=\frac{3}{4\pi^2}
\int^\infty_0\int^1_{-1}d|\vec{k}|d\cos\theta\vec{k}^2v_{\rm rel} \sigma^{\rm unpol}f(\vec{k}),
\label{eq:nvsigma}
\end{eqnarray}
where $n$ is the $\phi$ number density, $v_{\rm rel}$ is the relative velocity of the $\phi$ meson and 
the charmonium,
$\theta$ is the angle between the $\phi$ momentum $\vec{k}$ and the charmonium momentum and $f(\vec{k})$
is the Bose-Einstein distribution that $\phi$ mesons obey.
The thermal average of $v_{\rm rel} \sigma^{\rm unpol}$ is indicated by $\langle v_{\rm rel} \sigma^{\rm
unpol} \rangle$ \cite{NPA02Xu}.

\section{Numerical results and discussions}

\begin{figure}[htbp]
\centering
\includegraphics[width=15.0cm]{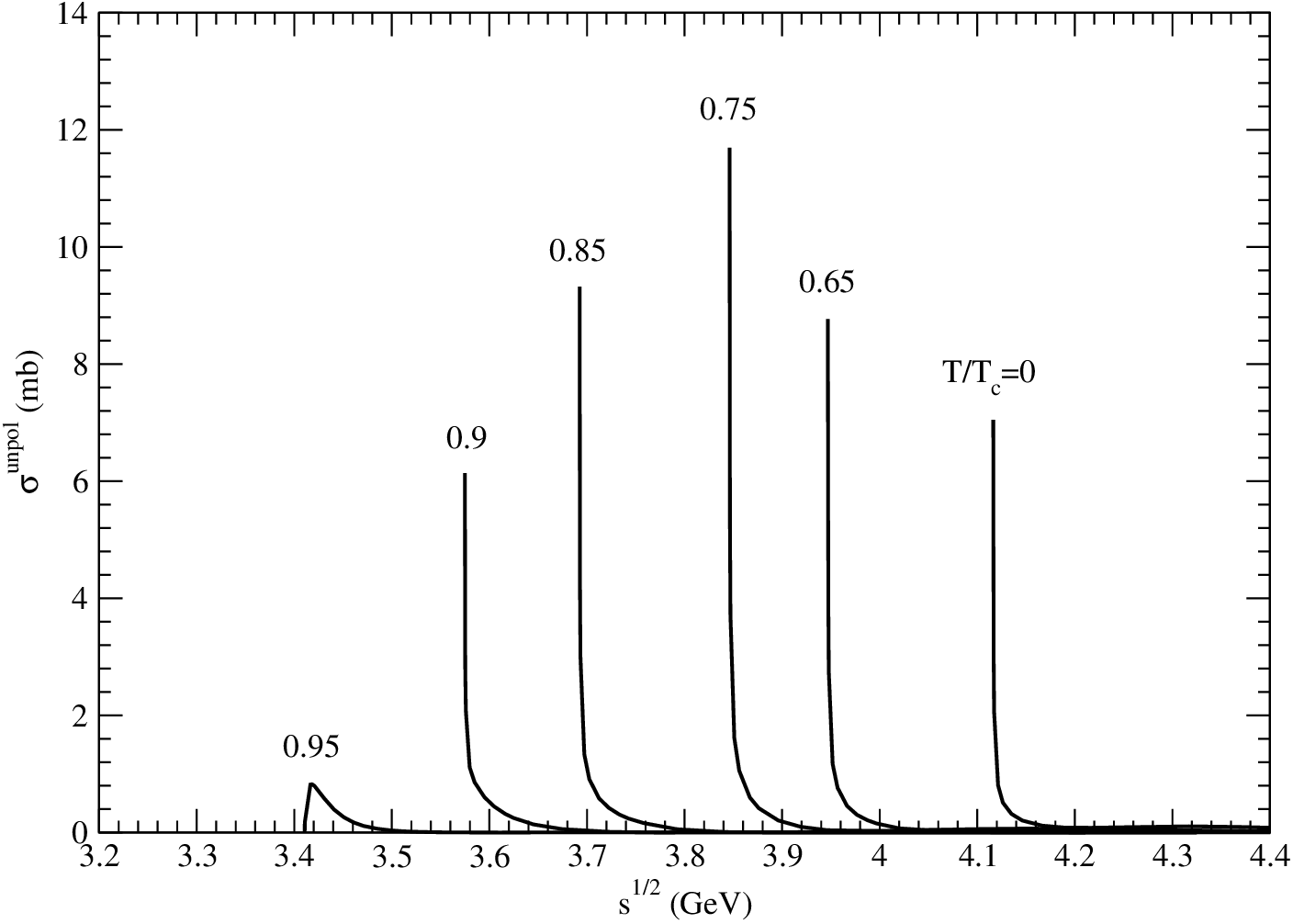}
\caption{Cross sections for $\phi J/\psi\to D^-_s D^+_s$ at
various temperatures.}
\label{fig:pjdsds}
\end{figure}

\begin{figure}[htbp]
\centering
\includegraphics[width=15.0cm]{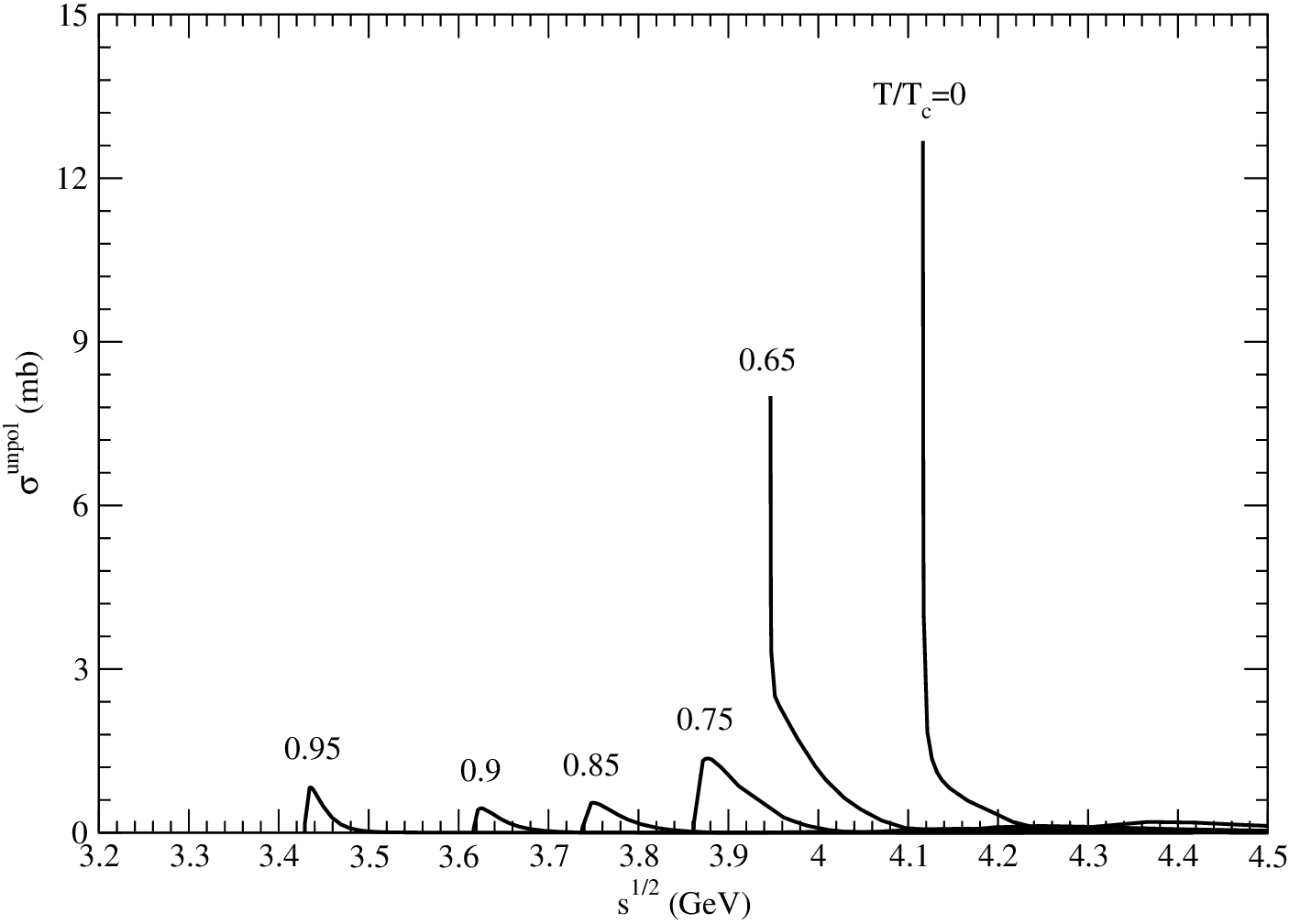}
\caption{Cross sections for $\phi J/\psi\to D^{*-}_s D^+_s$ or $D^-_s D^{*+}_s$ at
various temperatures.}
\label{fig:pjdsads}
\end{figure}

\begin{figure}[htbp]
\centering
\includegraphics[width=15.0cm]{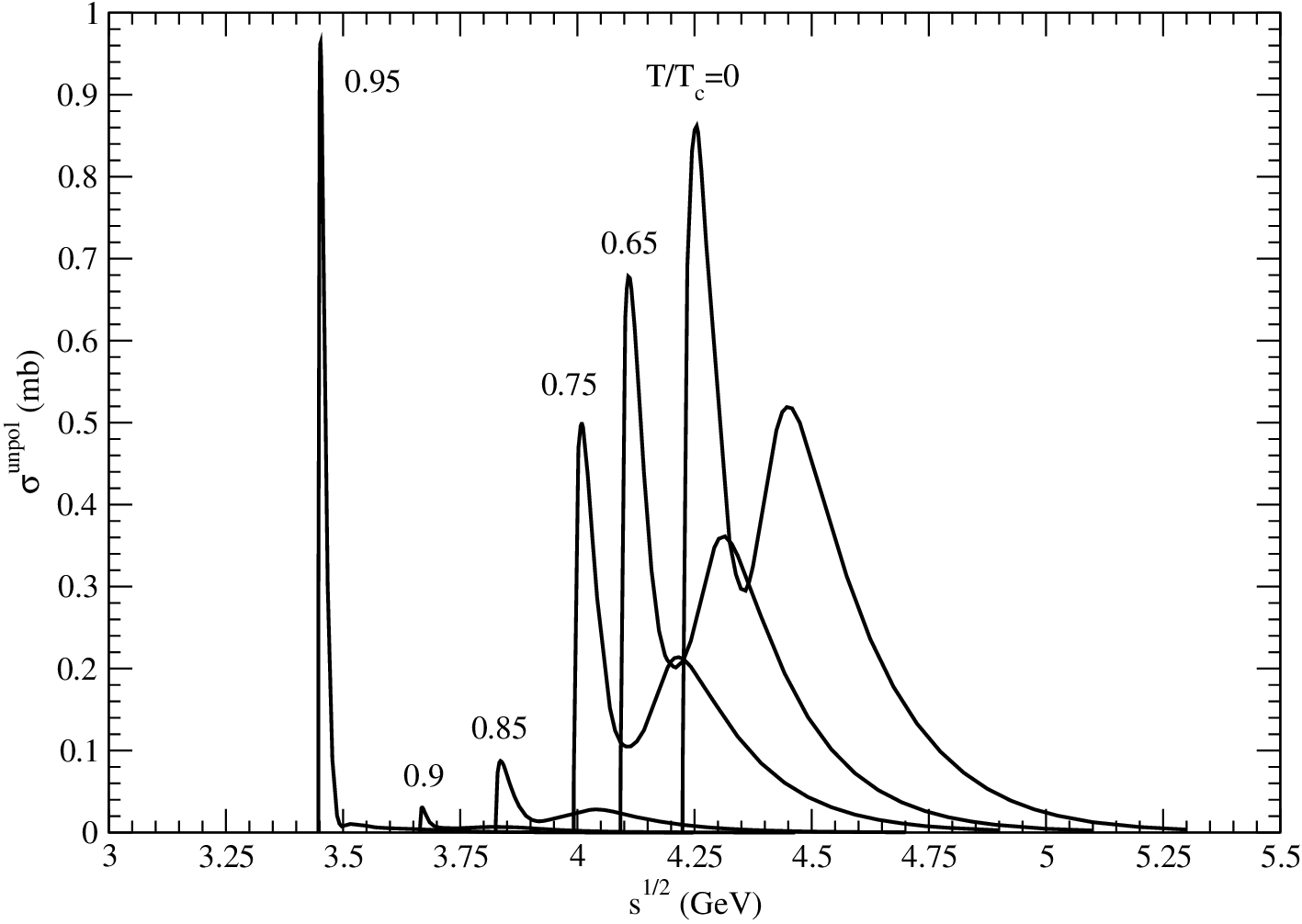}
\caption{Cross sections for $\phi J/\psi\to D^{*-}_s D^{*+}_s$ at
various temperatures.}
\label{fig:pjdsadsa}
\end{figure}

\begin{figure}[htbp]
\centering
\includegraphics[width=15.0cm]{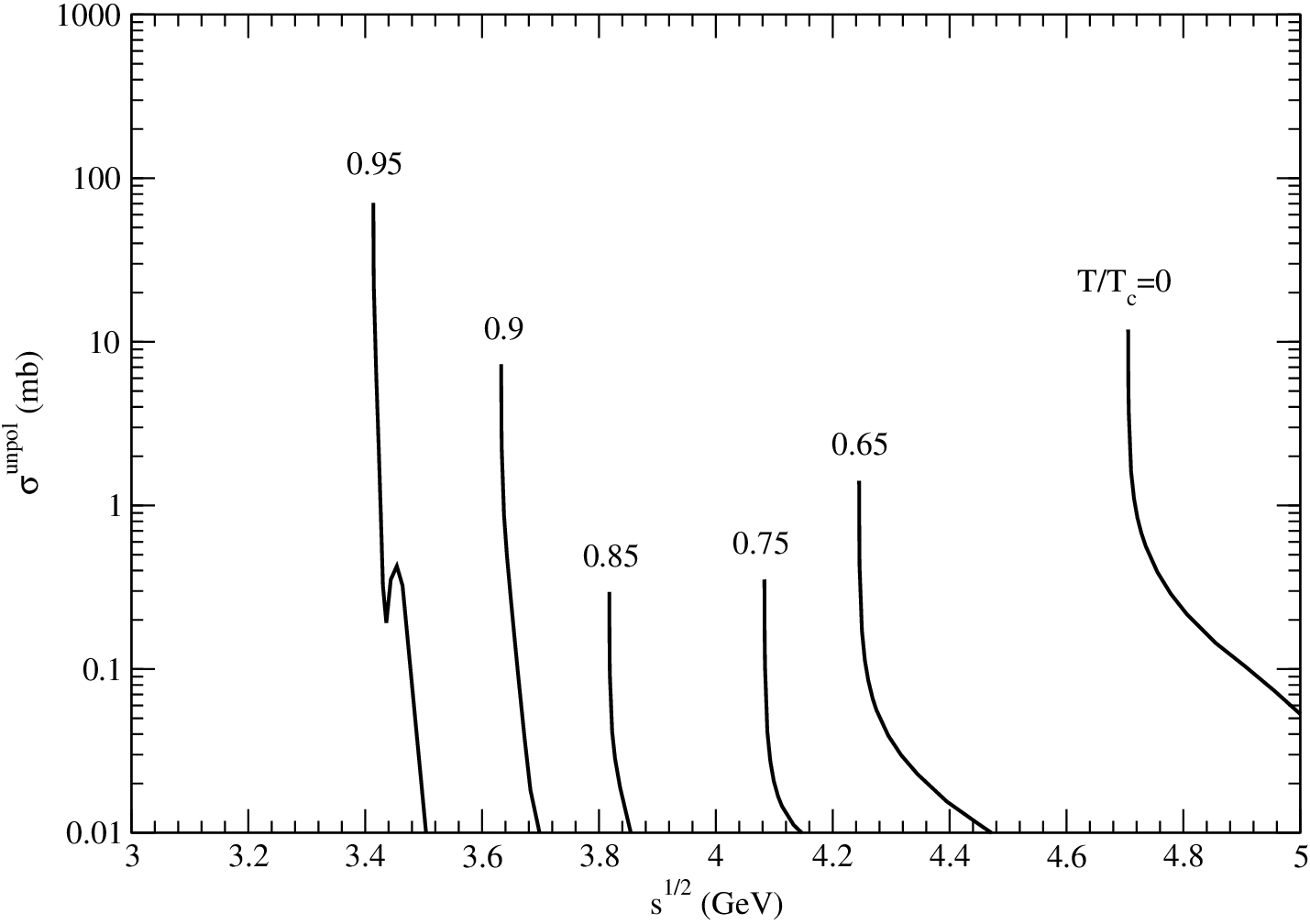}
\caption{Cross sections for $\phi \psi'\to D^-_s D^+_s$ at
various temperatures.}
\label{fig:ppdsds}
\end{figure}

\begin{figure}[htbp]
\centering
\includegraphics[width=15.0cm]{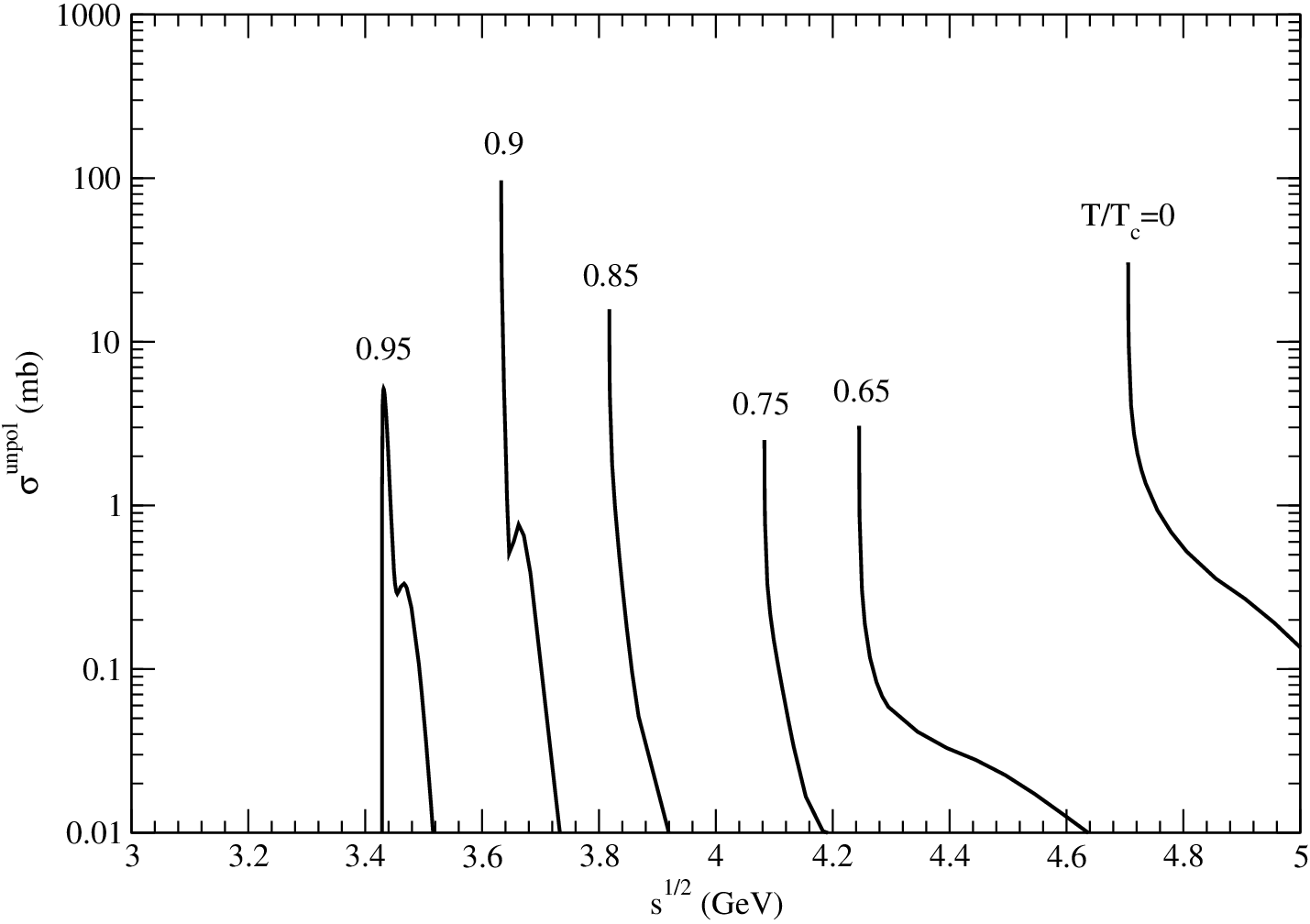}
\caption{Cross sections for $\phi \psi'\to D^{*-}_s D^+_s$ or $D^-_s D^{*+}_s$ at
various temperatures.}
\label{fig:ppdsads}
\end{figure}

\begin{figure}[htbp]
\centering
\includegraphics[width=15.0cm]{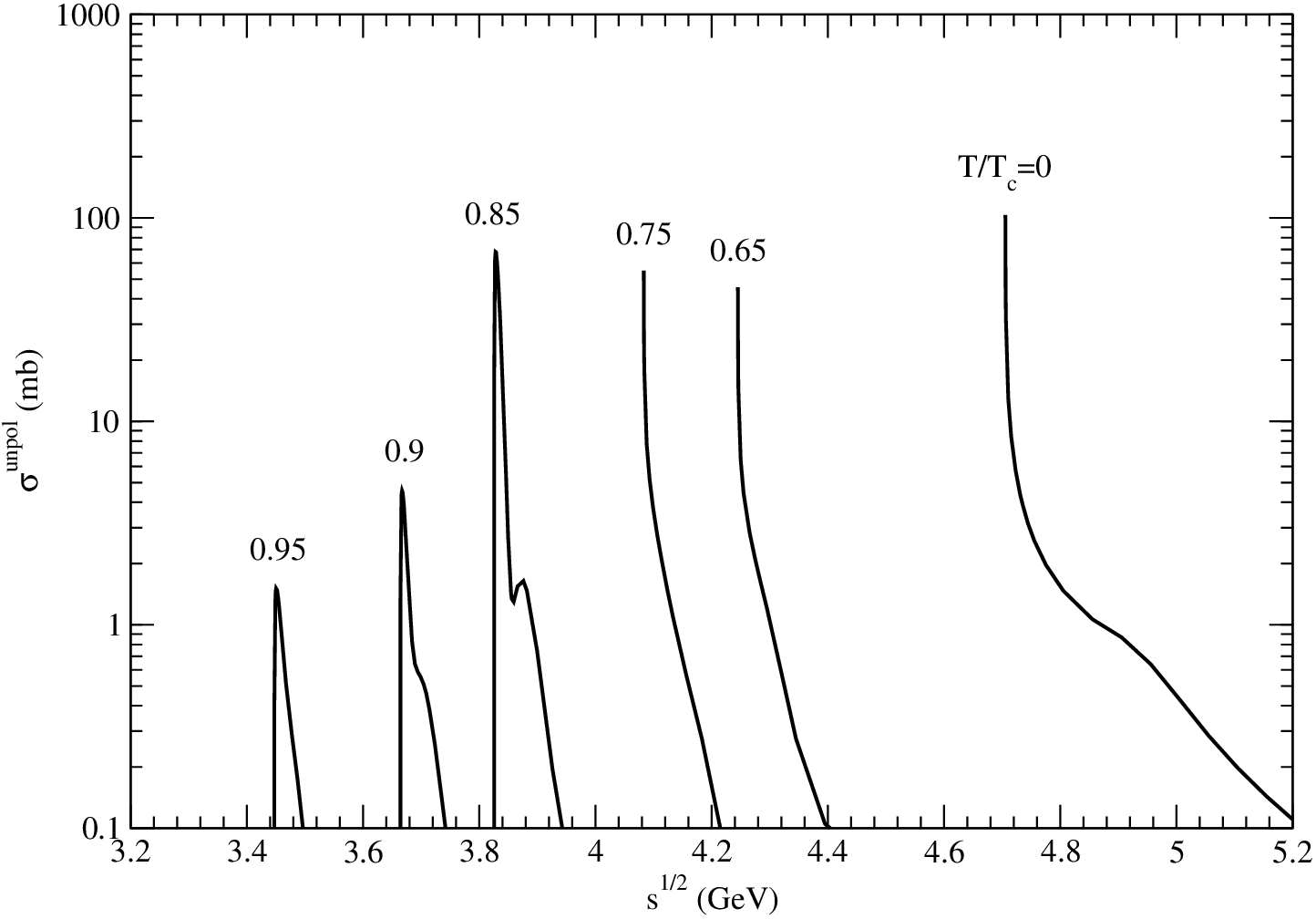}
\caption{Cross sections for $\phi \psi'\to D^{*-}_s D^{*+}_s$ at
various temperatures.}
\label{fig:ppdsadsa}
\end{figure}

\begin{figure}[htbp]
\centering
\includegraphics[width=15.0cm]{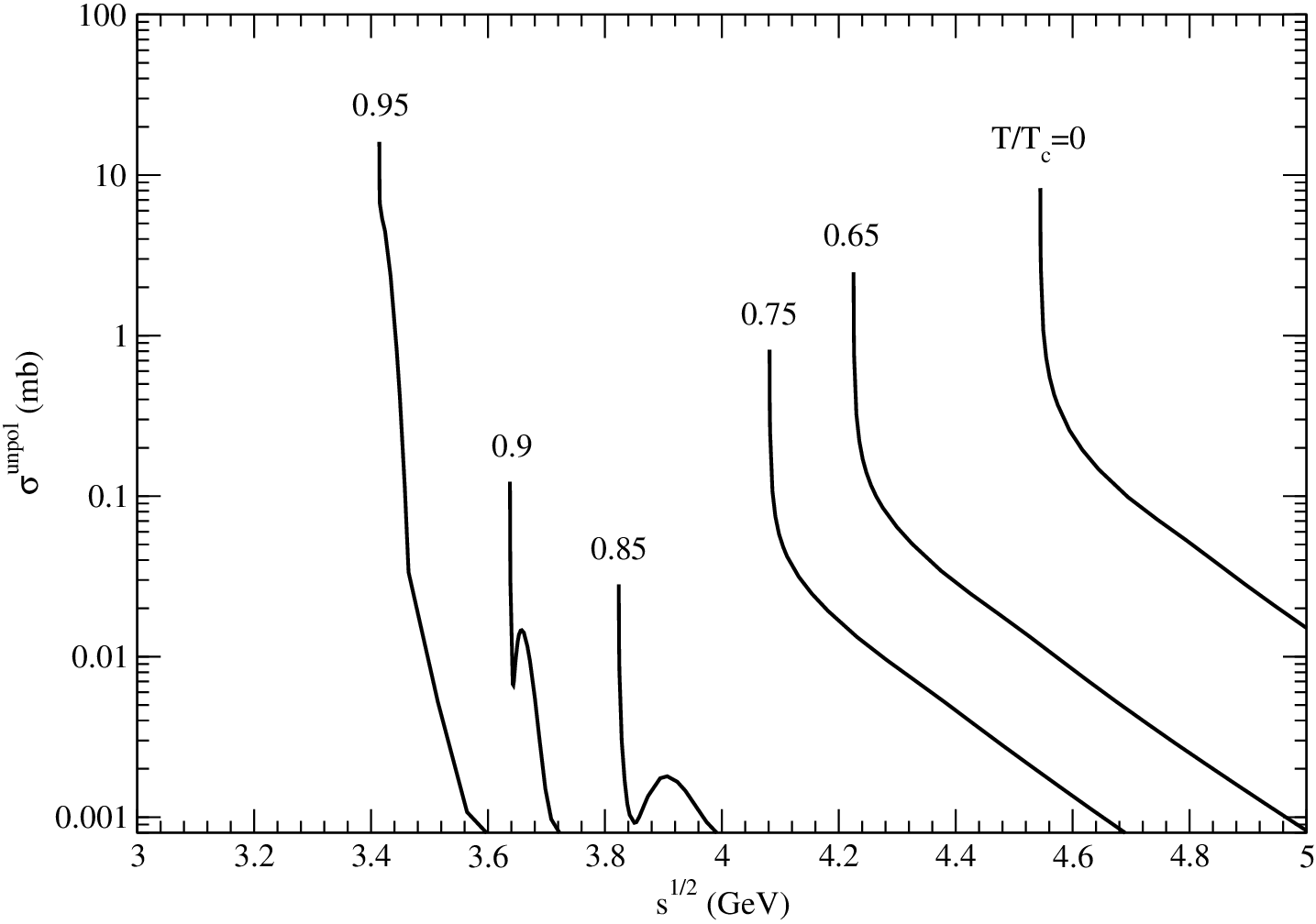}
\caption{Cross sections for $\phi \chi_{c}\to D^-_s D^+_s$ at
various temperatures.}
\label{fig:pcdsds}
\end{figure}

\begin{figure}[htbp]
\centering
\includegraphics[width=15.0cm]{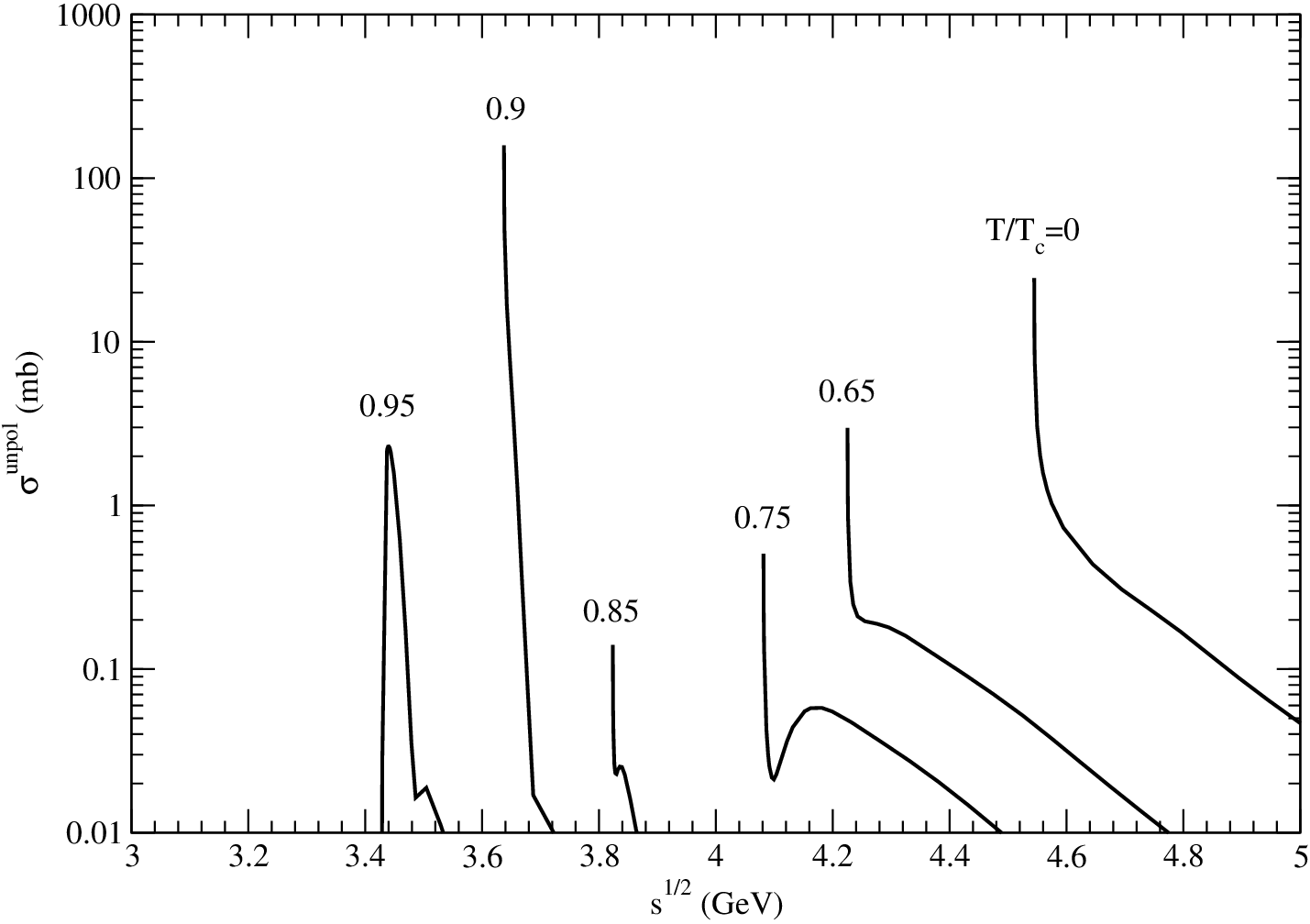}
\caption{Cross sections for $\phi \chi_{c}\to D^{*-}_s D^+_s$ or $D^-_s D^{*+}_s$ at
various temperatures.}
\label{fig:pcdsads}
\end{figure}

\begin{figure}[htbp]
\centering
\includegraphics[width=15.0cm]{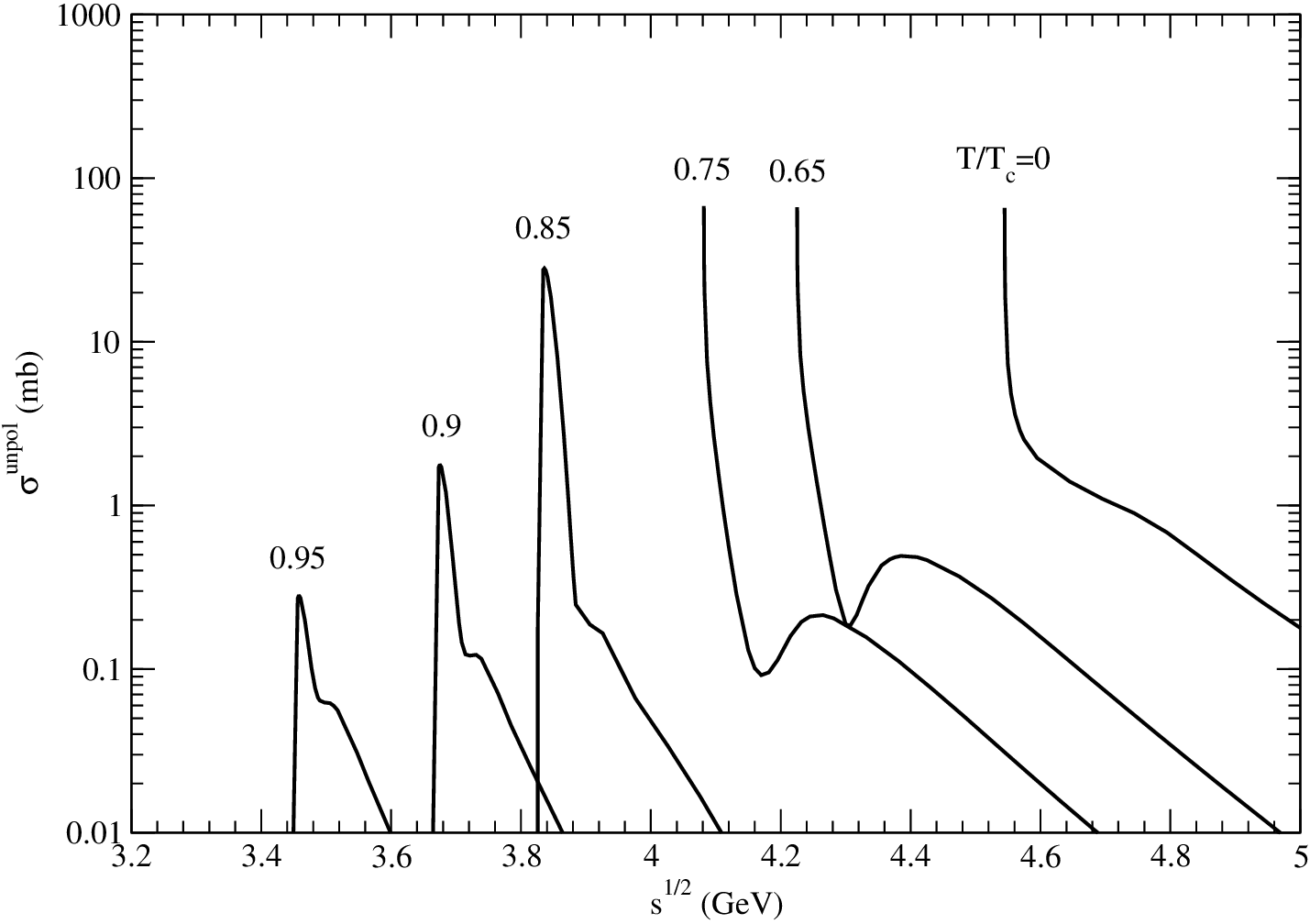}
\caption{Cross sections for $\phi \chi_{c}\to D^{*-}_s D^{*+}_s$ at
various temperatures.}
\label{fig:pcdsadsa}
\end{figure}

When a $\phi$ meson collides with one of $J/\psi$, $\psi'$ and $\chi_c$, $D^-_s D^+_s$,
$D^{*-}_s D^+_s$, $D^-_s D^{*+}_s$ or $D^{*-}_s D^{*+}_s$ is produced. Since the cross
section for the production of $D^{*-}_s D^+_s$ equals the cross section for the production
of $D^-_s D^{*+}_s$, they are shown in the same figure.
In Figs.~\ref{fig:pjdsds}-\ref{fig:pcdsadsa} we plot cross
sections for the following dissociation reactions:
$\phi J/\psi\to D^-_s D^+_s$,
$\phi J/\psi\to D^{*-}_s D^+_s$ or $D^-_s D^{*+}_s$,
$\phi J/\psi\to D^{*-}_s D^{*+}_s$,
$\phi \psi'\to D^-_s D^+_s$,
$\phi \psi'\to D^{*-}_s D^+_s$ or $D^-_s D^{*+}_s$,
$\phi \psi'\to D^{*-}_s D^{*+}_s$,
$\phi \chi_{c}\to D^-_s D^+_s$,
$\phi \chi_{c}\to D^{*-}_s D^+_s$ or $D^-_s D^{*+}_s$
and $\phi \chi_{c}\to D^{*-}_s D^{*+}_s$.
Each curve in a figure corresponds to one of the six temperatures, 0, 0.65$T_{\rm c}$, 0.75$T_{\rm c}$, 
0.85$T_{\rm c}$, 0.9$T_{\rm c}$ and 0.95$T_{\rm c}$.

Cross sections for $A+B \to C+D$ depend on masses of mesons $A$, $B$, $C$ and $D$. When
temperature approaches the critical temperature $T_{\rm c}$, $D^{\pm}_s$ and $D^{*\pm}_s$ become
degenerate in mass. The sum of the $D^-_s$ and $D^+_s$ masses almost equals the sum of
the $D^{*-}_s$ and $D^+_s$ masses or the sum of the $D^{*-}_s$ and $D^{*+}_s$ masses at 
$T$ = 0.95$T_{\rm c}$. However, the cross sections for
$\phi J/\psi\to D^-_s D^+_s$, $\phi J/\psi\to D^{*-}_s D^+_s$ and $\phi J/\psi\to D^{*-}_s D^{*+}_s$
are still different as shown in Figs.~\ref{fig:pjdsds}-\ref{fig:pjdsadsa}. This is because the
spin matrix elements involved in the transition amplitude are different for the three reactions.

The two reactions $\phi J/\psi\to D^{*-}_s D^+_s$ and $\phi J/\psi\to D^{*-}_s D^{*+}_s$
are endothermic at $T/T_{\rm c}=$ 0.75, 0.85 and 0.9. The sum of the $D^{*-}_s$ and $D^+_s$ masses
is smaller than the sum of the $D^{*-}_s$ and $D^{*+}_s$ masses. The square root of the Mandelstam
variable $\sqrt{s_{\rm pa}}$ corresponding to the peak cross section of $\phi J/\psi\to D^{*-}_s D^+_s$
is smaller than $\sqrt{s_{\rm paa}}$ corresponding to the peak cross section of 
$\phi J/\psi\to D^{*-}_s D^{*+}_s$.
The $\phi$ momentum $\vec{P}$ of $\phi J/\psi\to D^{*-}_s D^{+}_s$ at $\sqrt{s_{\rm pa}}$
has magnitude smaller than the one of $\phi J/\psi\to D^{*-}_s D^{*+}_s$ at $\sqrt{s_{\rm paa}}$. 
Accordingly, the factor $\frac {1}{s|\vec{P}|}$
in Eqs. (1) and (2) are larger for $\phi J/\psi\to D^{*-}_s D^+_s$ than for $\phi J/\psi\to 
D^{*-}_s D^{*+}_s$. On the other hand, the relative momentum $\vec{p}_{ab}$ of constituents
$a$ and $b$ due to small $|\vec{P}|$ in $\phi J/\psi\to D^{*-}_s D^+_s$
has magnitude smaller than that in $\phi J/\psi\to D^{*-}_s D^{*+}_s$.
The mesonic quark-antiquark relative-motion wave function $\psi_{ab}(\vec{p}_{ab})$
in the transition amplitude for $\phi J/\psi\to D^{*-}_s D^+_s$ is larger than the wave function for
$\phi J/\psi\to D^{*-}_s D^{*+}_s$. Therefore, the peak cross section of
$\phi J/\psi\to D^{*-}_s D^+_s$ is larger than the peak cross section of $\phi J/\psi\to D^{*-}_s D^{*+}_s$
at $T/T_{\rm c}=$ 0.75, 0.85 and 0.9 as seen in Figs.~\ref{fig:pjdsads} and \ref{fig:pjdsadsa}.

The reactions $\phi \psi'\to D^-_s D^+_s$ and $\phi \chi_{c}\to D^-_s D^+_s$ are
exothermic below $T_{\rm c}$. The reactions $\phi J/\psi\to D^-_s D^+_s$,
$\phi \psi'\to D^{*-}_s D^+_s$ and $\phi \chi_{c}\to D^{*-}_s D^+_s$
are exothermic below 0.95$T_{\rm c}$. The threshold energy of every exothermic reaction
is $m_{\phi}+m_{c\bar{c}}$ where $m_{\phi}$ and $m_{c\bar{c}}$ are the $\phi$ mass and the charmoinum mass,
respectively. We start calculating cross sections for exothermic reactions
at $\sqrt{s}=m_{\phi}+m_{c\bar{c}}+10^{-4}$ GeV and the cross sections at the energies
correspond to the curve tops. Exothermic reactions take place as long as the two initial mesons overlap and
even though the two initial mesons are at rest. However, the sizes of the
initial and final mesons affect cross sections. In $\phi J/\psi\to D^-_s D^+_s$
the sizes of $J/\psi$, $D^-_s$ and $D^+_s$ mesons at $T/T_{\rm c}=$ 0, 0.65 and 0.75
are so small that influence of
confinement is small at such sizes. Increase of the $J/\psi$
size with increasing temperature causes the cross section at $m_{\phi}+m_{J/\psi}+10^{-4}$ GeV
to increase. From $T/T_{\rm c}=$ 0.75 to 0.9 the $D^-_s$ ($D^+_s$) size increases quickly and
confinement becomes important. Since the plateau of $V_{\rm {si}}(\vec {r})$ at large
distances lowers with increasing temperature, confinement of the quark and the
antiquark to form $D^-_s$ and $D^+_s$ mesons becomes weaker and weaker and at $m_{\phi}+m_{J/\psi}+10^{-4}$
GeV the cross section thus decreases. In the $\phi\psi'$ and $\phi\chi_c$ reactions
the sizes of $\phi$, $\psi'$ and $\chi_c$ mesons are not
small and the influence of confinement on combining a quark and an antiquark
to form a $D^-_s$, $D^+_s$, $D^{*-}_s$ or $D^{*+}_s$ meson is appreciable at such sizes. 
Weakening confinement reduces cross sections with increasing temperature. When temperature increases
from $T/T_{\rm c}=$ 0.75 to 0.9, the $\phi$, $\psi'$ and $\chi_c$ sizes increase rapidly
and cross sections can thus go up. The two factors make the cross section for
each of the four $\phi\psi'$ or $\phi\chi_c$ reaction channels at the threshold energy plus $10^{-4}$ GeV
decrease first and increase next, as shown in Figs.~\ref{fig:ppdsds}, \ref{fig:ppdsads}, \ref{fig:pcdsds}
and \ref{fig:pcdsads}, while temperature changes from 0 to 0.9$T_{\rm c}$.

Charmonium dissociation in collision with $\phi$ meson has four channels.
The unpolarised dissociation cross section in the dissociation-rate formula is
the sum of the unpolarised cross sections for the four channels.
In collision of $\phi$ with $J/\psi$ ($\psi'$, $\chi_c$) $\sigma^{\rm unpol}$ is the unpolarised
cross section for $\phi J/\psi\to D^-_s D^+_s + D^{*-}_s D^+_s + D^-_s D^{*+}_s + D^{*-}_s D^{*+}_s$
($\phi \psi'\to D^-_s D^+_s + D^{*-}_s D^+_s + D^-_s D^{*+}_s + D^{*-}_s D^{*+}_s$,
$\phi \chi_c\to D^-_s D^+_s + D^{*-}_s D^+_s + D^-_s D^{*+}_s + D^{*-}_s D^{*+}_s$).

\begin{figure}[htbp]
\centering
\includegraphics[width=15.0cm]{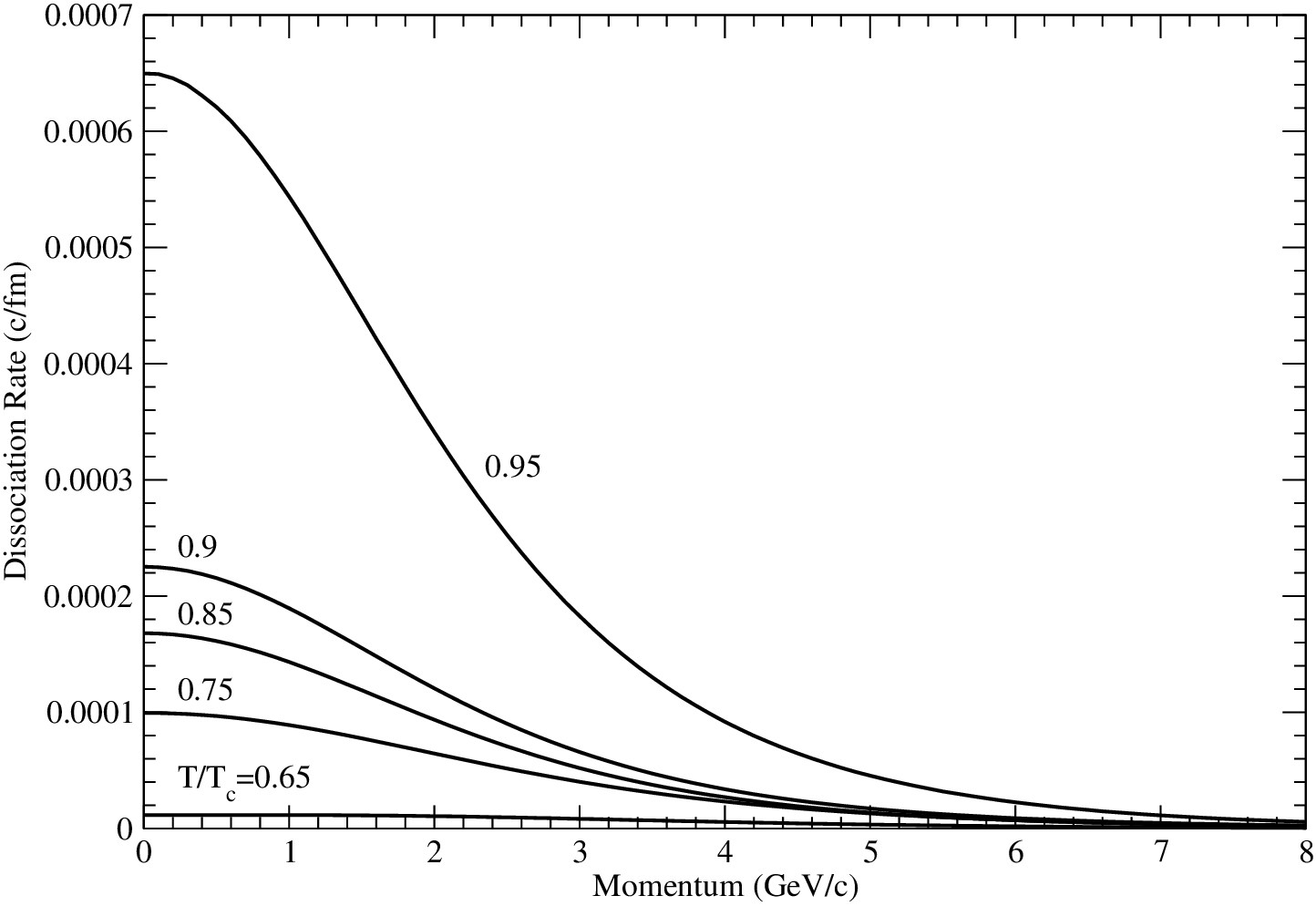}
\caption{Dissociation rate of $J/\psi$ with $\phi$ versus the $J/\psi$ momentum at various temperatures.}
\label{fig:pjnvs}
\end{figure}

\begin{figure}[htbp]
\centering
\includegraphics[width=15.0cm]{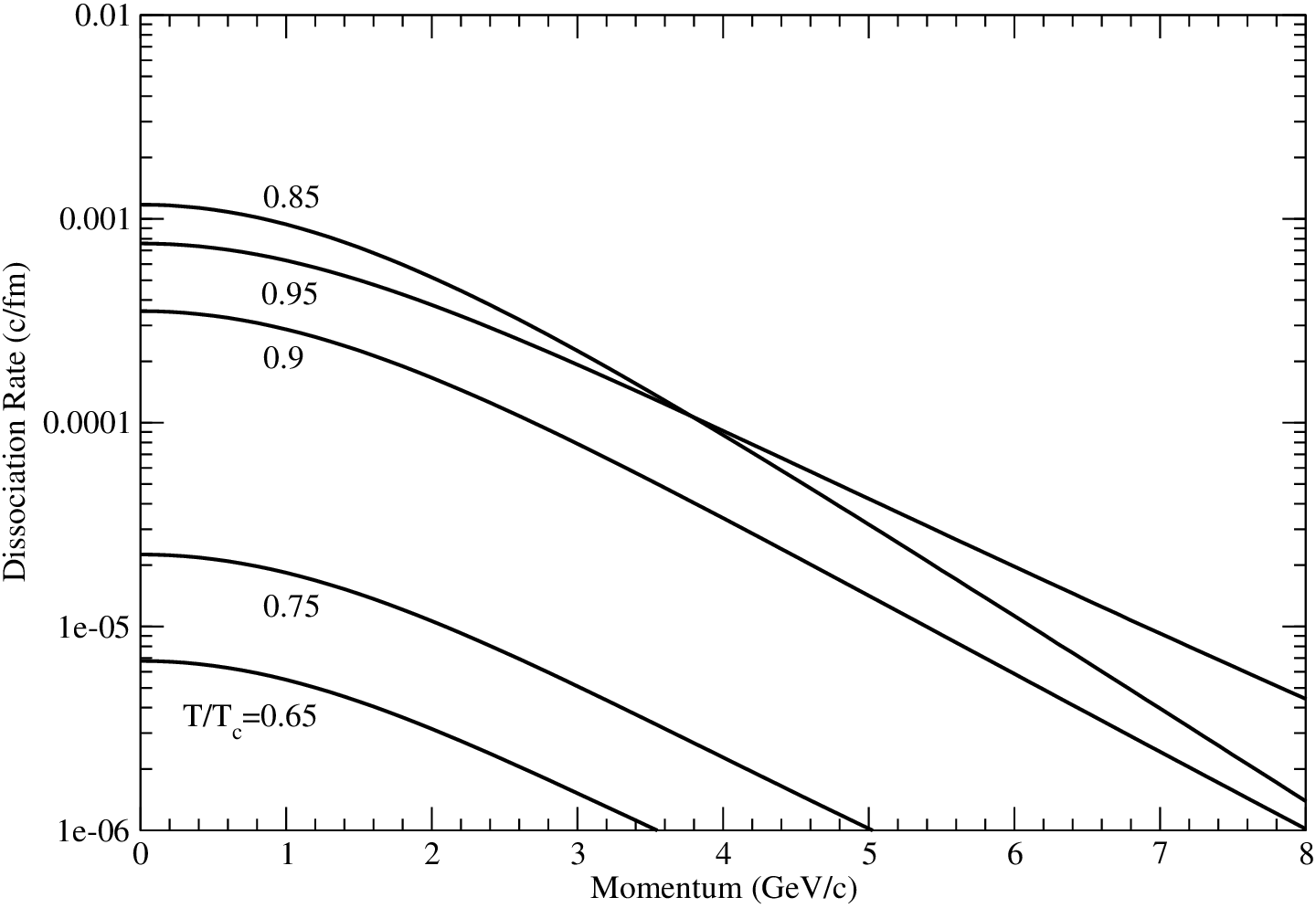}
\caption{Dissociation rate of $\psi'$ with $\phi$ versus the $\psi'$ momentum at various temperatures.}
\label{fig:ppnvs}
\end{figure}

\begin{figure}[htbp]
\centering
\includegraphics[width=15.0cm]{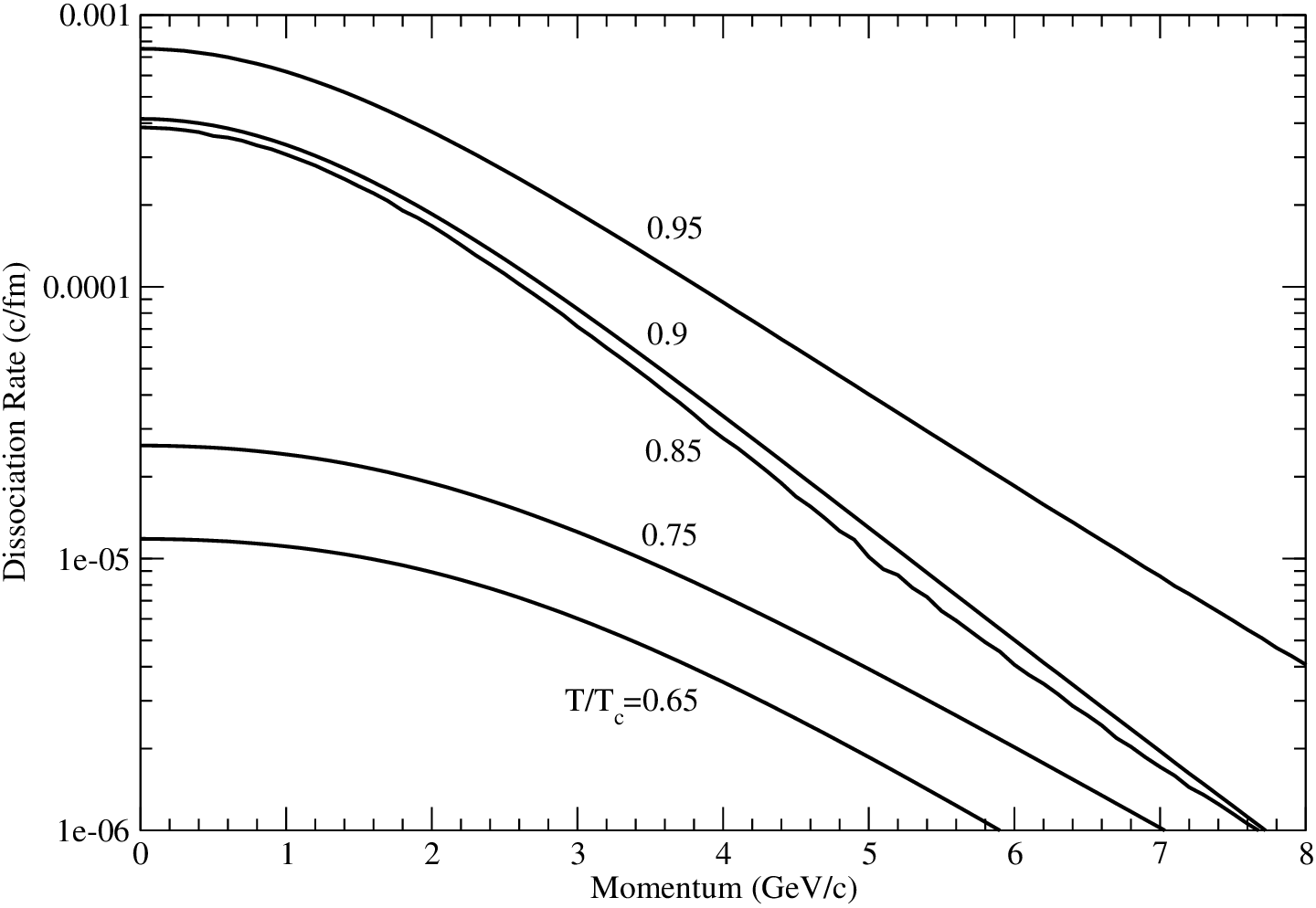}
\caption{Dissociation rate of $\chi_c$ with $\phi$ versus the $\chi_c$ momentum at various temperatures.}
\label{fig:pcnvs}
\end{figure}

Numerical results of dissociation rates as functions of charmonium momentum are plotted
in Figs. \ref{fig:pjnvs}-\ref{fig:pcnvs}. Since the $\phi$ distribution function, the unpolarised
cross section and the relative velocity in Eq. (10) vary with temperature, the dissociation rate depends on
temperature. As temperature increases from $T/T_{\rm c}=$ 0.65 to 0.95,
the $\phi$ mass decreases and the $\phi$ distribution function increases. This is a factor that
increases the charmonium dissociation rate. At zero charmonium momentum the slope of each curve is
zero. Because $\phi$ mesons obey the Bose-Einstein distribution, there is a domain of $\vec{k}$
in Eq. (10), which keeps $\sqrt{s}$ near the threshold energy so that the unpolarised cross section
contributes significantly. While the charmonium momentum increases, the
domain shrinks and the dissociation rate thus decreases. Since the dissociation rates of $J/\psi$,
$\psi'$ and $\chi_c$ with $\phi$ are less than 0.00066, 0.002 and 0.0008, respectively,
the charmonium suppression caused by the $\phi$-charmonium dissociation is weak.

\section{Summary}

Using the temperature-dependent quark potential that is derived from perturbative QCD
at short distances and the lattice gauge results at intermediate- and large-distances,
we obtain the unpolarised cross sections in the quark-interchange mechanism
and in the Born approximation. The reactions that we concern are
$\phi J/\psi\to D^-_s D^+_s$,
$\phi J/\psi\to D^{*-}_s D^+_s$ or $D^-_s D^{*+}_s$,
$\phi J/\psi\to D^{*-}_s D^{*+}_s$,
$\phi \psi'\to D^-_s D^+_s$,
$\phi \psi'\to D^{*-}_s D^+_s$ or $D^-_s D^{*+}_s$,
$\phi \psi'\to D^{*-}_s D^{*+}_s$,
$\phi \chi_{c}\to D^-_s D^+_s$,
$\phi \chi_{c}\to D^{*-}_s D^+_s$ or $D^-_s D^{*+}_s$
and $\phi \chi_{c}\to D^{*-}_s D^{*+}_s$.
The meson masses, confinement and mesonic quark-antiquark relative-motion wave functions cause
the large variation of the cross sections with respect to temperature. Medium effects on the dissociation
reactions are prominent. 
When the confinement gives similar contributions to two reaction channels, the spin-spin
interaction makes the difference of the unpolarised cross sections for the two channels. 
Using the unpolarised cross sections for the 12 reactions, we obtain the
dissociation rates of charmonia in the interactions with the $\phi$ meson. The rates generally
increase with increasing temperature or with increasing charmonium momentum. The rates are
quite small.

\begin{acknowledgments}
This work was supported by the National Natural Science Foundation
of China under Grant No. 11175111.
\end{acknowledgments}

\newpage

\newpage

\newpage

\newpage

\newpage

\newpage

\newpage

\newpage

\newpage

\newpage

\newpage

\newpage


\begin{references}
\bibitem{PRC98Matinyan} S. G. Matinyan and B. M\"{u}ller, Phys. Rev. C {\bf 58}, 2994 (1998).
K. L. Haglin, Phys. Rev. C {\bf 61}, 031902 (2000).
K. L. Haglin and C. Gale, Phys. Rev. C {\bf 63}, 065201 (2001).
Z. Lin and C. M. Ko, Phys. Rev. C {\bf 62}, 034903 (2000); J. Phys. G {\bf 27}, 617 (2001).
Y. S. Oh, T. S. Song and S. H. Lee, Phys. Rev. C {\bf 63}, 034901 (2001).
F. S. Navarra, M. Nielsen and M. R. Robilotta, Phys. Rev. C {\bf 64}, 021901(R) (2001).
L. Maiani, F. Piccinini, A. D. Polosa and V. Riquer, Nucl. Phys. A {\bf 741}, 273 (2004).
A. Bourque and C. Gale, Phys. Rev. C {\bf 78}, 035206 (2008); Phys. Rev. C {\bf 80}, 015204 (2009).
\bibitem{NPB79Peskin} M. E. Peskin, Nucl. Phys. B {\bf 156}, 365 (1979). G. Bhanot and M. E. Peskin, Nucl. 
Phys. B {\bf 156}, 391 (1979).
D. Kharzeev and H. Satz, Phys. Lett. B {\bf 334}, 155 (1994).
F. Arleo, P. B. Gossiaux, T. Gousset and J. Aichelin, Phys. Rev. D {\bf 65}, 014005 (2001).
\bibitem{PRC95Martins} K. Martins, D. Blaschke and E. Quack, Phys. Rev. C {\bf 51}, 2723 (1995).
C.-Y. Wong, E. S. Swanson and T. Barnes, Phys. Rev. C {\bf 62}, 045201 (2000); Phys. Rev. C {\bf 65}, 
014903 (2001).
T. Barnes, E. S. Swanson, C.-Y. Wong and X.-M. Xu, Phys. Rev. C {\bf 68}, 014903 (2003).
J. P. Hilbert, N. Black, T. Barnes and E. S. Swanson, Phys. Rev. C {\bf 75}, 064907 (2007).
\bibitem{PRC12Zhou} J. Zhou and X.-M. Xu, Phys. Rev. C {\bf 85}, 064904 (2012).
\bibitem{JPG15Ji} S.-T. Ji, Z.-Y. Shen and X.-M. Xu, J. Phys. G {\bf 42}, 095110 (2015).
\bibitem{PRC02Wong} C.-Y. Wong, Phys. Rev. C {\bf 65}, 034902 (2002).
\bibitem{PRD92Barnes} T. Barnes and E. S. Swanson, Phys. Rev. D {\bf 46}, 131 (1992). E. S. Swanson, Ann. 
Phys. (N.Y.) {\bf 220}, 73 (1992).
\bibitem{NPA07Li} Y.-Q. Li and X.-M. Xu, Nucl. Phys. A {\bf 794}, 210 (2007).
\bibitem{PRD81Buchmuller} W. Buchm\"{u}ller and S.-H. H. Tye, Phys. Rev. D {\bf 24}, 132 (1981).
\bibitem{NPB01Karsch} F. Karsch, E. Laermann and A. Peikert, Nucl. Phys. B {\bf 605}, 579 (2001).
\bibitem{PRD85Godfrey} S. Godfrey and N. Isgur, Phys. Rev. D {\bf 32}, 189 (1985).
\bibitem{NPA02Xu} X.-M. Xu, Nucl. Phys. A {\bf 697}, 825 (2002).
\bibitem{CPC14Olive} K. A. Olive {\it et al.}, Particle Data Group, Chin. Phys. C {\bf 38}, 090001 (2014).
\bibitem{PRD71Colton} E. Colton {\it et al.}, Phys. Rev. D {\bf 3}, 2028 (1971).
N. B. Durusoy {\it et al.}, Phys. Lett. B {\bf 45}, 517 (1973). 
W. Hoogland {\it et al.}, Nucl. Phys. B {\bf 126}, 109 (1977).
M. J. Losty {\it et al.}, Nucl. Phys. B {\bf 69}, 185 (1974).
\bibitem{Clarendon65Mott} N. F. Mott and H. S. W. Massey, {\em The Theory of Atomic Collisions} (Clarendon 
Press, Oxford, 1965).
T. Barnes, N. Black and E. S. Swanson, Phys. Rev. C {\bf 63}, 025204 (2001).
C.-Y. Wong and H. W. Crater, Phys. Rev. C {\bf 63}, 044907 (2001).
\end{references}
\end{document}